\documentclass[showpacs,preprint,amssymb]{revtex4}
\usepackage{graphicx}
\usepackage{dcolumn}
\usepackage{bm}
\def\be{\begin{equation}}
\def\ee{\end{equation}}
\def\bea{\begin{eqnarray}}
\def\eea{\end{eqnarray}}
\def\line{\hbox to \hsize}
\def\frac #1#2{{#1\over #2}}

\def\tr{{\rm  tr\,}}
\def\det{{\rm det\,}}
\def\hpsi{\hat \psi}
\def\psid{\psi^{\dagger}}
\def\hpsid{\hat \psi^{\dagger}}

\def\ad{a^{\dagger}}
\def\bd{b^{\dagger}}

\def\ad{ a^{\dagger}}

\def\det{{\rm det\,}}

\def \ket #1{{\vert #1\rangle}}

\def \brak #1#2{{\langle#1\vert#2\rangle}}
\def\eval #1#2#3{{\langle#1\vert#2\vert#3\rangle}}

\def\1{\mbox{\bf 1}}

\begin{document}

\title{Fusion rules and  vortices in $p_x+ip_y$ superconductors
}

\author{ MICHAEL STONE}

\affiliation{University of Illinois, Department of Physics\\ 1110 W. Green St.\\
Urbana, IL 61801 USA\\E-mail: m-stone5@uiuc.edu}

\author{SUK-BUM CHUNG}

\affiliation{University of Illinois, Department of Physics\\ 1110 W. Green St.\\
Urbana, IL 61801 USA\\E-mail: sukchung@uiuc.edu}

\begin{abstract}
The ``half-quantum" vortices ($\sigma$) and quasiparticles ($\psi$)
in a two-dimensional $p_x+ip_y$ superconductor obey the Ising-like
fusion rules $\psi\times \psi=1$, $\sigma\times \psi=\sigma$, and
$\sigma\times \sigma= 1+\psi$. We explain how  the physical fusion
of vortex-antivortex pairs allows us to use these rules   to read
out the information encoded in the topologically protected space of
degenerate ground states. We comment  on the potential applicability
of this fact to quantum computation.

\end{abstract}

\pacs{ 71.10.Pm, 73.43.-f, 74.90.+n}

\maketitle

\section{Introduction}

The magic  of a quantum computer\cite{benioff,manin,feynman,shor94}
is that it  makes a virtue out of a vice. The numerical simulation
of even simple quantum systems requires exponentially large storage
space and exponentially  long computation times. The quantum state
of a  collection   of $N$ spin-$\textstyle{1\over 2}$ particles, or
qubits, requires $2^N$   classical variables for its description,
and   solving the Schr{\"o}dinger equation to follow   its  time
evolution  requires  a conventional computer to perform up to
$2^N\times 2^N$ operations per time step.  Conversely, a  computer
built out of  quantum components  would   be able to    exploit  the
massive parallelism inherent in  quantum time evolution to provide
fast  solutions for  problems that  would require exponential time on
conventional machines.

The  difficulties   to be overcome  in  building a quantum computer
are many.  The quantum  system at its core  must be strongly isolated
from the  environment so as to avoid decoherence;  the unitary
transformations that constitute the    elementary computational steps
must  be performed with sufficient precision that error-correcting codes
\cite{shor96} remain effective;  and, after all  the quantum computation
has been performed,  the  system must be capable of being reconnected to
the outside world in such a way  that  the output   can be  read off
{\it via\/} a  measurement process that does not disturb the result.

A   particularly appealing   scheme for quantum computation exploits
topologically protected macroscopic quantum states in many-body
systems \cite{freedman_AMS,kitaev_annals,bravyi,sarma} . The first
examples of such  topologically ordered and protected many-body
quantum states were found by Xiao-Gang  Wen and Qian Niu
\cite{wen-niu} in the context of the fractional quantum Hall effect.
Wen and Niu observed that if we place a filling-fraction $\nu=1/n$
quantum Hall state on a genus $g$ Riemann surface (a thought
experiment only!) then there are  $n^g$ degenerate, and essentially
indistinguishable, ground states. They  also observed that by
creating a vortex antivortex pair in the Hall fluid and then moving
the vortex   around one of the homology generators of the surface
they  could cause the system to roll over from one degenerate state
to the next. Since the degeneracy is lifted only by the
exponentially suppressed tunneling of vortices around the
generators,  the coherence  of  a linear superposition of such
states  is {\it topologically protected\/}. Furthermore, as
described above, such superpositions  can in principle  be
manipulated by controlling   the motion of their vortex  defects,
and so  inducing  on them  representations of the braid group of
vortex world-lines.

For simple systems, such as the $\nu=1/n$ Hall state on  a
simply-connected surface, these braid-group representations are
abelian,   interchange and braiding giving rise only to a phase
factor. This is  not sufficient  for    a universal quantum computer
\cite{bravyi,freedman_modular}. We need a  non-abelian
representation---{\it i.e.\/} particles  with non-abelian statistics
\cite{moore-read}. There are a various  two-dimensional physical
systems that should, in theory, have excitations with
non-abelian statistics. Those closest to being  realized in practice
are   all associated with many-body wavefunctions that contain a
Pfaffian factor. These  candidates are:   i) a quantum Hall effect
state seen at filling-fraction $\nu=5/2$ \cite{read-green}, ii)   an
anticipated, but as yet unobserved, phase of  a rapidly rotating
atomic Bose gas, iii) a    two-dimensional $p_x+ip_y$
superconducting state---an example of which  has possibly  been
observed  in the layered Sr$_2$RuO$_4$ superconductor
\cite{rice95,baskaran96}.   While the non-abelian statistics of
these Pfaffian states is still not quite adequate for the
construction of a universal computer, it is still worthy of study.

The mathematical origin of the non-abelian braiding in the first two
candidate systems is quite technical,  involving  complicated linear
dependences among various   Pfaffian  wavefunctions
\cite{nayak-wilczek}. The  non-abelian braid statistics  of  the
$p_x+ip_y$ superconductor  is much easier to understand  physically.
The mechanism  has been beautifully explained   by Ivanov
\cite{ivanov} and  further clarified   by Stern {\it et al.\/
}\cite{stern}.

The essential ingredient  is that the core of an Abrikosov vortex in
any  superconductor  contains localized low-energy bound states
whose existence is guaranteed by an index theorem  that relates
the phase winding number $n$ of the vortex to the number of branches
of low energy excitations \cite{volovik_index}.  The  $p_x+ip_y$
superconductor  is  unusual in that  it can host a
``half-quantum''   vortex \cite{volovik_book}, where the phase
winding is seen by only one of the two components of the electron
spin. The quasiparticles bound in the core of a two-dimensional
version of such a  vortex  are  Majorana,  {\it i.e.\/} they are
identical to their antiparticles. Further, for  odd phase winding
numbers,  each vortex binds   a  zero energy mode  whose
field-expansion coefficient is hermitian. In the presence of  $2N$
such vortices, the many-body Hilbert  space of  zero-energy states
is   $2^{N-1}$ dimensional. This number  is  in  contrast with   the
na{\"\i}ve count of $2^{2N}$ dimensions that  would come from
(wrongly) supposing  that each single-particle zero-energy mode must
be either occupied or unoccupied.  Because these $2^{N-1}$
independent       states cannot be associated with any individual
vortex,   the  many-body   state  encodes highly non-local
information. Ivanov \cite{ivanov} demonstrated  that the operation
of braiding or  interchanging  the vortices acts on this
zero-energy  space  in the same way that braiding acts on the
$2^{N-1}$ dimensional space of  $N$-point  conformal blocks  in  the
level   $k=2$, ${\rm SU}(2)$ Wess-Zumino-Witten (WZW) model.

It is not surprising  that this particular WZW model plays some
role. It  is known that it describes the low energy physics of
Pfaffian quantum Hall state \cite{fradkin-nayak-tsvelik-wilkzek}. It
also describes the low energy physics of the Bose gas pfaffian state
\cite{ardonne-kedem-stone,fradkin-nayak-schoutens}, and  is   a
component of the low energy effective action of the  two-dimensional
$p_x+ip_y$ superconductor \cite{stone-roy}.

Because the zero-energy  states consist of an equal superposition of
particle and hole, they are electrically neutral. They are also
nonlocal. As a consequence they  should be little affected by local
impurities and external fields, and so be  topologically protected.
The vortices binding them, however, carry magnetic flux and so might
manipulable by external probes such as STM tips.  These  vortices
are thus potential candidates for the basic building blocks  of a
quantum computer.  The   problem is that the internal state in the
$2^{N-1}$-dimensional protected zero-energy  space  is so decoupled
from the outside world that it is hard to see how  it can be
measured.

This read-out problem is not unique to non-abelian braid statistics.
There is as yet no completely convincing experimental demonstration
of   even the   abelian ``anyonic'' statistics in the $\nu=1/n$ Hall
state---although no theorist doubts that it is present.  An
experiment that can detect non-abelian statistics is even  harder
to perform. We need some  method of accessing  the non-abelian state
by  reading out the information in  the protected space.

One approach, at least in principle, is to use  the ``fusion rules''
of the vortices  to re-connect the protected states  to  something
measurable. The vortex  fusion  algebra contains, in addition to the
obvious rule that phase-winding number adds,  a $\mathbb{Z}_2$
factor  that is  essentially that   of the Ising model primary-field
operators  $\mathbb{I}$, $\sigma$,    and $\psi$
\cite{big_yellow_book}, or of the analogous affine primary fields
(of spin zero, one-half, and one respectively)  in the  ${\rm
SU}(2)_2$ WZW model: \be \psi\times \psi=\mathbb{I} ,\quad
\sigma\times \sigma= \mathbb{I}+\psi,\quad \psi\times\sigma= \sigma.
\label{EQ:fusion_algebra} \ee In the superconductor  it is natural
to identify  $\mathbb{I}$ with the ground state and $\psi$ with the
Bogoliubov quasiparticle.   The latter  is conserved only mod 2
because  two  quasiparticles can combine to form a Cooper pair and
vanish  into the condensate. Hence $\psi\times \psi\to \mathbb{I}$.
Slightly less obvious is the  identification of    $\sigma$  with
the odd-winding-number vortex.  This comes about  because a pair of
vortices is associated with a single zero-energy  state that can be
occupied ($\psi$) or unoccupied ($\mathbb I$). Hence $\sigma\times
\sigma = \mathbb{I}+\psi$.  The third product, $\sigma\times
\psi=\sigma$, is then determined by   the  associativity of the
fusion-product algebra. The principal claim of this paper is that it
is possible to use this algebra to determine some of the information
in the topologically protected space   by physically fusing  a
vortex with an antivortex  and seeing if they leave behind a real
quasiparticle $\psi$.

In subsequent sections, we will fill in the details of this process.
In section two we discuss the form of the zero energy solutions to
the single-particle Bogoliubov-de-Gennes  equation, focusing on the
sign ambiguities which are the ultimate source of the
particle-creation legerdemain.   In section three we show  how
tunneling between nearby vortices can be used to resolve the sign
ambiguity, and how  braiding  followed by fusing reveals details of
the protected state.  Section four is a conclusion and discussion.
An appendix describes some necessary, but  technical, aspects of
Berry transport  in a superconducting state.

\section{Vortex core states}

The topological objects  that have the properties we seek to exploit
are the ``half-quantum'' vortices  of a   thin film of $p_x+ip_y$
spin-triplet superfluid, such as $^3$He-A, or possibly  the
superconductor Sr$_2$RuO$_4$. The order parameter in such a
superfluid contains an angular-moment vector ${\bf l}$, a vector
${\bf d}$ characterizing the Cooper-pair spin state, and an overall
phase-factor  $\exp(i\chi)$.   In a half-quantum vortex  the vector
${\bf l}$ lies along the vortex axis and the vector ${\bf d}$ lies
in plane which, for convenience, we imagine  to be perpendicular to
${\bf l}$. As we encircle the vortex the phase $\chi$ increases by
$\pi$ while  the ${\bf d}$ vector rotates  through $180^\circ$.
Although $\exp(i\chi)$ and ${\bf d}$ separately  change sign, it is
their product that appears in the order parameter, and this is
single-valued. The effect of the combined rotation is that the
spin-up fermions see the  order parameter  phase wind through
$2\pi$, while the spin-down fermions see no phase change  at all.
Andreev reflection off the winding phase will bind  low energy
spin-up quasiparticles modes in the vortex core, but there will be
no spin-down quasiparticle states with energy significantly  less
that the gap $|\Delta|$.   Because only one component of spin is
involved in the low energy physics, we will  from now on   regard
the fermions as being spinless.  It is   this spinlessness  that
allows  the $p_x+ip_y$ superconductor to have topological properties
analogous to those of the  spin polarized  Pfaffian quantum Hall
state \cite{read-green}.

Consider a two-dimensional film of  superfluid with its ${\bf l}$
vector perpendicular to the film, and with a winding-number $n$
vortex at the origin.  We are principally interested in the case
$n=\pm1$, but we will keep $n$ general for the moment, so as to
bring out  a  distinction  between vortices with  $n$ even and $n$
odd. The  Bogoliubov-de-Gennes (BdG)   equation \footnote{For a
discussion of the BdG equation for a  $p_x+ip_y$ superfluid see, for
example, \cite{stone-roy}. } has  bound-state solutions  of the form
\be
\left(\matrix{u\cr v}\right)= e^{il\theta} \left(\matrix{u_l\cr
v_l}\right)
\label{EQ:vortex_bs1}
\ee
where
\be
\left(\matrix{u_l\cr
v_l}\right)= \left(\matrix{ e^{i\theta(n+1)/2}[a(r)
H_{l+1/2}(k_fr)+{\rm c.c.}]\cr e^{-i\theta(n+1)/2}[b(r)
H_{l-1/2}(k_fr)+{\rm c.c}]}\right).
\label{EQ:vortex_bs2}
\ee
Here
$H_l(k_fr)$ is a Hankel function of the radial co-ordinate $r$, the
angle  $\theta$ is the polar co-ordinate, and $k_f$ is the Fermi
momentum. The coefficient functions $a(r)$, $b(r)$  are  slowly
varying on the length scale of $k_f^{-1}$ and are most conveniently
written in terms of a pair of auxiliary variables $x$ and
$\theta(x)$. These   are defined in terms of the impact parameter
$r_0\equiv l/k_f$ by
\bea
x&=&\sqrt{r^2-r_0^2},\\
\theta(x)&=&  \tan^{-1}\left(\frac x{r_0} \right).
\eea
If we  write
\be
\left(\matrix{a(r)\cr b(r)}\right) =
\left(\matrix{e^{-in\theta(x)/2} \tilde a(x)\cr e^{in\theta(x)/2}\tilde b(x)}\right)
\ee
then, in the  Andreev approximation, the coefficients   $\tilde a(x)$ and $\tilde b(x)$ obey
\be
\left(\matrix{-iv_f \partial_x& \Delta(r) e^{in\theta(x)}\cr
                     \Delta(r)e^{-in\theta(x)} &+iv_f\partial_x}\right)\left(\matrix{\tilde a(x)\cr \tilde b(x)}\right)=
                     E_l \left(\matrix{\tilde a(x)\cr \tilde b(x)}\right).
\label{EQ:chordal_equation} \ee
The one-dimensional  eigenvalue
equation (\ref{EQ:chordal_equation}) has a physically appealing
interpretation  as describing the propagation of a quasiparticle
along a rectilinear  trajectory  with $r_0$ its distance of closest
approach to   the vortex centre. The variable   $x$ is the distance
along the trajectory  with $x=0$ being the point of closest
approach. As the quasiparticle   moves it sees  the   local value of
the order parameter  $\Delta(r(x))\exp[ in  \theta(x)]$. The Hankel
function is  evanescent for $r<r_0$, and $\tilde a$ and $\tilde b$
can be taken as being constant in this region.  In order for the
Hankel functions to combine  to give the Bessel functions
$J_{l\pm1/2}(kr)$ that remain finite at $r=0$, we need to impose the
condition that  $a$ and $b$ (and hence $\tilde a$ and $\tilde b$) be
real at $x=0$. This condition allows us to  extend the definition of
$\tilde a(x)$ and $\tilde b(x)$  to  continuous functions on  the
entire real line  by setting  $\tilde a(-x)=(\tilde a(x))^*$ and
$\tilde b(-x)=(\tilde b(x))^*$. It may be verified  that $(\tilde
a(x),\tilde b(x))^T$ continues to satisfy
(\ref{EQ:chordal_equation})   in the extended domain. The
one-dimensional equation  is able to capture the physics of the
bound states because the Andreev scattering  that confines the
particle/hole  in the vortex core is almost exactly retro-reflective
\cite{stone-vortex}.

Solving the one-dimensional problem shows that, for small $l$, the
energy eigenvalue is given by $E(l)=  - l\omega_0$ where $\omega_0$
is a frequency determined by the order parameter profile
$\Delta(r)$.  This frequency is positive if $n>0$ and negative if
$n<0$. A  glance  at (\ref{EQ:vortex_bs1}) and (\ref{EQ:vortex_bs2})
shows that  single-valuedness of the BdG wavefunction requires $l$
to take integer values when $n$ is odd, and half-integer values when
$n$ is even.  The spectrum may  therefore be labeled by an integer
$m$ and is given by
\be
E_m=\cases{ - \omega_0 m, & $n$ odd\cr
                       -\omega_0(m+1/2), & $n$ even.}
\ee
Both cases are consistent with the $E\leftrightarrow -E$
symmetry of the BdG eigenvalues. For  odd winding numbers the
spectrum  contains an $E=0$ mode. If we create $N$ vortices there
will be one  of these  zero modes localized  in each vortex core.
There will, however, always be   an even number of zero modes. For
an  odd number of vortices,  an additional  zero-mode will be found
on the boundary of the superfluid \cite{stone-roy}.

Although the zero modes for  the $n=+1$ and $n=-1$  vortices both
have  $l=0$, their actual angular dependence  differs. For the
$n=+1$ vortex, whose circulation is in the same sense as the
$p_x+ip_y$  Cooper-pair angular momentum, we have
\be
\left(\matrix{u_0(r,\theta) \cr v_0(r,\theta)}\right)=
\left(\matrix{ e^{i\theta} f(r)\cr e^{-i\theta}f(r)}\right)
\ee
For
the $n=-1$ vortex we have,
\be
\left(\matrix{u_0(r,\theta) \cr
v_0(r,\theta)}\right)= \left(\matrix{ f(r)\cr f(r)}\right).
\ee
In
both cases  the radial function $f(r)$ is real and decays as $\exp(
- \Delta r /v_{f})$   away from  the vortex core.

We have been tacitly   assuming  that $\Delta$ is real. A global
phase rotation $\Delta\to e^{i\chi} \Delta$ alters  each of the
eigenmodes and we  can choose an  overall phase for the modes so
that
\be
\left(\matrix{u_l(r,\theta) \cr v_l(r,\theta)}\right)\to
\left(\matrix{e^{i\chi/2}u_l(r,\theta) \cr
e^{-i\chi/2}v_l(r,\theta)}\right).
\label{EQ:gauge_choice}
\ee
An
inspection of the detailed form of the solutions shows that, with
this phase choice,  we retain the property that the upper component
of eigenmode $l$ is the complex conjugate of the lower component for
eigenmode $-l$.  This property does  not uniquely specify  the
$(u,v)^T$ vectors, however.  Even after normalization, so that $\int
(|u_l|^2+|v_l|^2)\,d^2x =1$,  a  choice of overall sign  remains to
be made.

In the vicinity of the vortex core we can use the bound states to
make a  mode expansion of the low energy part of the fermion field
\be
\left(\matrix{\hpsi(r,\theta)\cr\hpsid(r,\theta) }\right)=\sum_l
b_l\left(\matrix{u_l\cr v_l}\right )e^{il\theta}
+\hbox{higher-energy modes}.
\ee
The fact that the lower field
component $\hpsid$ is the hermitian conjugate of the upper component
$\hpsi$ coupled with the phase choice made in
(\ref{EQ:gauge_choice}),  enforces the hermiticity condition
$b_l=b_{-l}^\dagger$ on the annihilation and creation operator
coefficients. This property is characteristic of a Majorana fermion.
The  bound-state quasiparticles are therefore their own
antiparticles. For $n$ odd,  the  zero-energy, mode coefficient
obeys  $b_0=b_0^\dagger$ and cannot be thought of as being either a
creation or annihilation operator. Instead, the conditions
$\{\hpsi(x),\hpsi(y)\}=0$ and $\{\hpsid(x),\hpsi(y)\}=\delta^2(x-y)$
together with  the completeness of the eigenmodes tell us that
\be
b_0^2=1/2,\quad \{b_0,b_l\}=\{b_0,\bd_l\}=0, \, \,\, l\ne0.
\ee

We have taken the trouble in this section to  display the
bound-state wavefunctions in some detail. We did this  to make clear
the origin of the $b_l^\dagger=b_{-l}$ Majorana condition, to make
explicit  our   phase choices, and to point out  the remaining sign
ambiguity.  This  last point  is important. For   conventional
fermion annihilation and creation operators, the relative sign of
$a_i$ and $\ad_i$ is fixed by the condition $\{a,\ad\}=1$,  but the
relative sign of the various $a_i$ with respect to one another  is
unimportant  as it   can always be absorbed into the definition of
the free-field modes whose coefficients they are.  The relative sign
of the $b_0$'s associated with different vortices has a physical
significance that  will   be appreciated after we consider tunneling
between nearby vortex pairs.

\section{Braiding and Fusion}

Suppose we start from a homogeneous ($\Delta = {\rm const.}$) state
and adiabatically create $N$ vortex-antivortex pairs.  For ease of
discussion imagine the  the pairs arranged so that each  antivortex
lies  on the $x$ axis, and the corresponding vortex lies  vertically
above. We   label the pairs from left to right by the index
$i=1,\ldots,N$.   In the core of each vortex and antivortex there
will be a zero-mode with its corresponding $b_0$ mode operator.  Up
to a factor, these hermitian operators obey gamma-matrix
anti-commutation relations,  so it is natural to  rename these
$b_0$'s   as $\gamma_i/\sqrt{2}$   (vortex) and and
$\gamma_{i+N}/\sqrt{2}$ (antivortex). These $2N$ operators  can be
assembled to make $N$ each of annihilation and creation operators
\bea
b_i&=&\frac 12(\gamma_i+i\gamma_{i+N}),\nonumber\\
\bd_i&=&\frac 12 (\gamma_i-i\gamma_{i+N}).
\eea
These  $b_i$ and
$\bd_i$'s    act on a   $2^N$-dimensional space of degenerate ground
states. This  space is split  into two physically equivalent
$2^{N-1}$-dimensional spaces by a superselection rule: although
fermion number conservation  is broken by the presence of the
condensate, no physical process internal to the system can change an
odd total fermion number into an even fermion number.  Which of the
two spaces we are in depends on whether the total number of
particles in the system is  odd or even.

The odd-even decomposition is reflected in the fusion algebra. By
using the rules (\ref{EQ:fusion_algebra}) to repeatedly fuse a
$\sigma$  with $\sigma\times \sigma= \mathbb{I}+\psi$, we find
\bea
\sigma\times\sigma\times \sigma &=& \sigma+\sigma,\nonumber\\
\sigma\times\sigma\times \sigma\times \sigma &=& \mathbb{I} +\psi+\mathbb{I}+\psi,\nonumber
\eea
{\it etc.\/}
In general, for $2N$ vortices,
\be
\underbrace{\sigma\times\cdots\times \sigma}_{  2N\,{\rm factors}}
 = (\underbrace{\mathbb{I}+\cdots+\mathbb{I}}_{2^{N-1}\,{\rm terms}}) + (\underbrace{ \psi+\cdots +\psi}_{2^{N-1}\,{\rm terms}}).
\ee
The  $2^{N-1}$-dimensional multiplicity spaces of the $\mathbb
I$ and the $\psi$ are separately  invariant under the action of the
braid group.

When each vortex ``$i$'' is brought close to its sibling antivortex
``$N+i$,'' the $2^{N}$-fold ground state degeneracy is lifted.
Because two nearby odd-winding number vortices are effectively a
vortex with  an even winding number, tunneling between  their two
zero-energy BdG modes  will cause them to realign  and  become two
eigenmodes  with  some small non-zero energy $\pm E_i$. In the
second quantized Bogoliubov  Hamiltonian  these two single-particle
modes  make  one state  that can either be occupied   and contribute
$E_i$ to the total energy or be unoccupied and contribute energy
$-E_i$.   If we were to slowly merge each vortex with its sibling
antivortex,  the $\pm E_i$ eigenmodes will split further  and
eventually merge with  the continuum of unbound states.  If the
$E_i$ state is empty, nothing remains after the vortices annihilate.
If it is occupied, an unbound    quasiparticle with energy of
$|\Delta|$ is left behind.

It is not easy to compute the matrix elements  for the tunnel
splitting, but we do not need to know them  in detail. When the
splitting is small we can restrict ourselves to the  $2^{N}$
dimensional space of nearly degenerate states, and in  this space
the many-body Hamiltonian must be of the form
\bea
\hat H_0&=& \sum_{i=1}^N E_i\frac 1{4i} [\gamma_{N+i},\gamma_i]
+\frac 12 \sum_{i=1}^N E_i\nonumber\\
&=& \frac 12 \sum_{i=1}^N E_\alpha (\bd_i b_i-b_i\bd_i) +\frac 12 \sum_{i=1}^N E_i\nonumber\\
&=& \sum_{i=1}^N E_i \bd_i b_i.
\label{EQ:H_0}
 \eea
This is because  $\frac 1{4i} [\gamma_{N+i},\gamma_i]$ is the only
hermitian  operator that can be made solely out of $\gamma_i$ and
$\gamma_{i+N}$, and so the only possible operator that tunneling
between vortex ``$i$'' and ``$N+i$'' can contribute to  the
Hamiltonian. Hermiticity demands that   that   the $E_i$ be real. We
can also assume that the $E_i$ are   positive by using this
requirement  to fix the relative  sign ambiguity between
$\gamma_{i+N}$ and $\gamma_i$.

The lowest energy state $\ket{0}$  for an even number of particles
is that  which is annihilated by all the $b_i$.  In this ground
state all fermions are paired, and all of the  not-quite-zero-energy
states are empty. If the system contains an odd number of particles,
however, one  will  be left unpaired, and will   occupy the  lowest
of the  not-quite-zero-energy  states.

Now we consider how we can manipulate the occupation numbers of the
not-quite-zero modes. Following Ivanov \cite{ivanov}, we
adiabatically transport vortex $i$  around vortex $j$ and bring it
back to its original position. In this process the local phase
$\chi$ seen by each vortex will increment by $\pm 2\pi$, and so
cause the phase factors $e^{ i\chi/2}$ (see
(\ref{EQ:gauge_choice}))  in   the  zero modes of  vortex $i$ and
$j$ to  change sign.  The field operators  $\hat \psi(x)$,
$\hat\psid(x)$  are indifferent to the choice of modes in which we
expand them, and must be  unchanged by the braiding process. The
sign change of the mode vector  $(u_0,v_0)$ must therefore be
compensated by  a change in   the sign of the mode coefficients
$\gamma_i$ and $\gamma_j$.   Consequently
\vbox{
$$
\hat H_0  =\cdots  + E_i\frac 1{4i} [\gamma_{N+i},\gamma_i]+\cdots + E_j\frac 1{4i} [\gamma_{N+j},\gamma_j]
+\cdots+\frac 12 E_i +\cdots+ \frac 12 E_j+\cdots  \nonumber
$$
\centerline{is changed to}
$$
\hat H_0^{{\rm new}} =\cdots  - E_i\frac 1{4i} [\gamma_{N+i},\gamma_i]+\cdots - E_j\frac 1{4i} [\gamma_{N+j},\gamma_j]
+\cdots+ \frac 12 E_i +\cdots + \frac 12 E_j+\cdots
$$
}
In addition to this explicit monodromy in $\hat H_0$, the state
$\ket{0}$  might acquire a  non-abelian, holonomy  from   Berry
transport.  It was argued by Stern {\it et al.\/} \cite{stern} ,
that, for the choice of phases in  $(u_0,v_0)$, the state  $\ket{0}$
is at most multiplied by an overall Berry phase, and so  essentially
returns to itself after the braiding process.   We agree with this
conclusion, although we find the discussion in \cite{stern}
unnecessarily involved. We therefore provide our own derivation of
this key fact in the appendix.

After the braiding  the state  $\ket{0}$ remains an  eigenvector of
all the  $\frac 1{4i} [\gamma_{N+k},\gamma_k]$   with eigenvalue
$-1/2$.  Because of the sign changes in $\hat H_0$,  however, it is
no longer the lowest energy state of $\hat H_0^{\rm new}$. It is
instead an excited state with energy $E_i+E_j$, corresponding to the
two not-quite-zero modes of the $i$ and $j$-th pairs being occupied.
If we  now  adiabatically fuse  the $i$-th vortex with its sibling
antivortex  we recover the  uniform  state together  with a
quasiparticle $\psi$.   The same is true for for the $j$-th pair.

\begin{figure}
\includegraphics[width=4.0in]{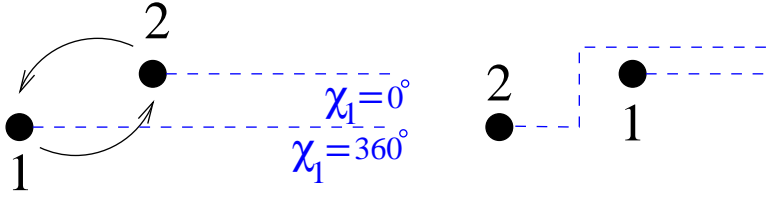}
\caption{
\label{braid}\sl Phase changes due to the braid operation
$T_1$: Before the braiding the phase $\chi_2(1)$ seen by vortex 1
due to vortex 2 is just more than  $180^\circ$, and the phase
$\chi_1(2)$ seen by vortex 2 due to vortex 1 is just more than
$0^\circ$. After the braiding the phase $\chi_1(2)$ at vortex 2 due
to vortex 1 is just more than  $180^\circ$,  but because vortex 1
remains below  below the displaced vortex-2 branch cut, the phase
$\chi_2(1)$   is now just more than $360^\circ$. Consequently
$\gamma_1$ is replaced by $\gamma_2$, but $\gamma_2$ is replaced by
$-\gamma_1$.
}
\end{figure}

We now suppose we  make an anti-clockwise interchange $T_i$ of the
$i$ and $i+1$-th vortices.   In order to follow what happens, we
must first fix the sign ambiguity between the  $\gamma_i$. We
therefore draw ``branch cuts'' to the right of each vortex and
parallel to the $x$-axis and set the $\chi_i$ produced by the vortex
equal to zero immediately above the cut.  The local phase $\chi(i)$
seen by  vortex $i$ is then the sum of the $\sum_{j\ne i} \chi_j$'s
of the other vortices and anti-vortices. Under the interchange, and
keeping track of the explicit monodromy  of the local $\chi_i$'s, we
find that the zero mode of the $i+1$-th vortex  replaces  that of
the $i$-th while the $i$-th zero mode replaces  {\it minus\/}  that
of the $i+1$-th.  Thus (see fig \ref{braid} and \cite{ivanov})
\be
T_i :\cases{\gamma_i\to\phantom{-}\gamma_{i+1}&
\cr \gamma_{i+1}\to
-\gamma_i.&}
\ee

Since the geometric arrangement of the vortices is unchanged, the
Hamiltonian becomes
$$
\hat H_0^{\rm new} =\cdots  + E_i\frac 1{4i} [\gamma_{N+i},\gamma_{i+1}]+
\cdots - E_{i+1}\frac 1{4i} [\gamma_{N+i+1},\gamma_i]
+\cdots+ \frac 12 E_i +\cdots + \frac 12 E_{i+1}+\cdots.
$$
The crucial effect  is not  so much the sign change but  that the
Berry transport of   $\ket{0}$  preserves its property that it is
the  state killed by the $b_k$ of the  {\it original\/}  $\frac
1{4i} [\gamma_{N+k},\gamma_k] $. Consequently $\ket{0}$  is  no
longer an eigenstate of the new Hamiltonian,  but is instead a
linear superposition of eigenstates. The outcome  of fusing vortex
$i$ with its antivortex is no longer certain. Same is true for the
outcome of fusing vortex $i+1$ with its antivortex. In fact, it will
be shown later that we have constructed an entangled state.

The sign change in $\gamma_i\to -\gamma_j$ does have significance,
however, as it ensures that  the result of  braiding     is  to take
$\gamma_i\to {\gamma}'_i$ with
$$
{\gamma}'_i={\gamma}_jO_{ji}=U{\gamma}_iU^{-1},
$$
where $O_{ij}$ is an element of  ${\rm SO}(2N)$, as opposed to ${\rm
O}(2N)$. Here  $U$ is a spin-representation   matrix  that  would
act on the  $2^N$ dimensional space of degenerate ground states were
we  to make the braid group act by holonomy on the states, instead
of  by  explicit monodromy on  the Hamiltonian. The ${\rm O}(2N)$
spin representation is irreducible, but under restriction  to
${\rm SO}(2N)$   it decomposes into  two irreducible components,
these  being the spaces of odd or even fermion number. For the
elementary braiding operation $T_i$, Ivanov showed  \cite{ivanov}
that the relevant unitary operator $U(T_i)$ can be taken to be \be
 \tau_i= \frac{1}{\sqrt 2}(1+\gamma_{i+1}\gamma_{i}),
\ee
as this is unitary and obeys
\bea
\tau_i \gamma_i \tau_i^{-1}&=&  \gamma_{i+1},\nonumber\\
\tau_i \gamma_{i+1} \tau_i^{-1}&=&- \gamma_{i}.\nonumber
\eea
The
operation of taking vortex $i$ completely around vortex $i+1$ and
back to its starting point is therefore
\bea
\tau_i^2 \gamma_i \tau_i^{-2}&=& - \gamma_{i},\nonumber\\
\tau_i^2 \gamma_{i+1} \tau_i^{-2}&=&- \gamma_{i+1}.\nonumber
\eea
For the rest of this section we will take the holonomy point of
view---{\it i.e.\/}\ the Hamiltonian will be kept fixed as $\hat H_0
= \sum_i E_i \bd_ib_i$, and  a  braiding $T$ will act on the state
by $U(T)$.

We now  consider the effect of various  braid group generators  on
the occupation number basis  states
\be \ket{n_1,\ldots,n_N}\equiv
({\bd}_1)^{n_1}\cdots ({\bd}_N)^{n_N}\ket{0}.
\ee
When  we expand
out the generator $\tau_i$ in terms of the annihilation and creation
operators, we find that
\be
\tau_i= \frac{1}{\sqrt
2}(1+b_{i+1}b_i+\bd_{i+1}b_i+b_{i+1}\bd_i+\bd_{i+1}\bd_i),
\ee
and
so
\bea
\tau_i \ket{n_1,\ldots,n_i,n_{i+1},\ldots,n_N}&=&\frac
1{\sqrt 2} \left\{ \ket{n_1,\ldots,n_i,n_{i+1},\ldots,n_N}\right.
\nonumber\\
&&\qquad +\ket{n_1,\ldots,(n_i-1),(n_{i+1}-1),\ldots, n_N}\nonumber\\
&&\qquad +  \ket{n_1,\ldots,(n_i-1),(n_{i+1}+1),\ldots n_N}\nonumber\\
&&\qquad -  \ket{n_1,\ldots,(n_i+1),(n_{i+1}-1),\ldots,n_N}\nonumber\\
&&\qquad \left.-
\ket{n_1,\ldots,(n_i+1),(n_{i+1}+1),\ldots,n_N}\right\}.
\label{EQ:VVInter}
\eea
Here we understand that when the $\pm 1$
takes the occupation number $n_j$ out of the set $\{0,1\}$ the
illegal state is to be replaced by zero.

We next  define $T^{(0)}_i$ to be the operation  of interchanging
vortex $i$ with its sibling  antivortex. (A vortex and an antivortex
being distinguishable, this operation does not return the system to
its original configuration and hence cannot be considered as a
braiding operation.) The corresponding unitary operator is
\bea
\tau^{(0)}_i &=&\frac{1}{\sqrt 2} (1+\gamma_{i+N}\gamma_i)\nonumber\\
&=& \frac 1{\sqrt 2} (1+i(\bd_ib_i-b_i\bd_i)).
\eea
From this, we
find
\be
\tau^{(0)}_i \ket{n_1,\ldots, n_i,\ldots,n_N} = e^{i\pi
n_i/4} \ket{n_1,\ldots, n_i,\ldots,n_N}.
\ee
One can also consider
an operation involving interchanging vortex $i$ with the antivortex
of its partner, $i+1$, to the right. We will call this operation
$T^{(1)}_i$. The corresponding operator is
\be
\tau_i^{(1)}=\frac
1{\sqrt{2}}(1-ib_{i+1}b_i +i\bd_{i+1}b_i -i
b_{i+1}\bd_i+i\bd_{i+1}\bd_i),
\ee
and this acts as
\bea
\tau_i^{(1)} \ket{n_1,\ldots,n_i,n_{i+1},\ldots,n_N}&=&\frac 1
{\sqrt 2}\left\{ \ket{n_1,\ldots,n_i,n_{i+1},\ldots,n_N}\right.
\nonumber\\
&&\qquad -i\ket{n_1,\ldots,(n_i-1),(n_{i+1}-1),\ldots, n_N}\nonumber\\
&&\qquad +i  \ket{n_1,\ldots,(n_i-1),(n_{i+1}+1),\ldots n_N}\nonumber\\
&&\qquad +i
\ket{n_1,\ldots,(n_i+1),(n_{i+1}-1),\ldots,n_N}\nonumber\\ &&
\left.\qquad
-i\ket{n_1,\ldots,(n_i+1),(n_{i+1}+1),\ldots,n_N}\right\}.
\label{EQ:VAInter}
\eea

A strategy for accessing the protected information is now apparent.
The wavefunction of the protected state $\ket{\Psi}$ belonging to
the even (or odd) fermion number sector is specified by the
$2^{N-1}$ complex numbers forming the its components in the
occupation-number basis $\ket{n_1,\ldots,n_N}$ . Provided that we
can have access to multiple  copies of $\ket {\Psi}$, we can
repeatedly  fuse the vortex-antivortex  pairs and so estimate  the
probability of a particular pattern $(n_1,\ldots n_N)$ of relict
particles. In this way we  obtain  the numbers
$|\brak{\Psi}{n_1,\ldots,n_N}|^2$. We lose all relative phase
information in this process, however.  This loss occurs not only
because we are finding probabilities, but also because  the
different occupation-number states have unpredictably different
tunneling energies, and so time evolution will scramble their
relative phases during the fusion process.

Not all is lost, however. (\ref{EQ:VVInter}) and (\ref{EQ:VAInter})
seem to suggest that we can use direct fusion combined with
controlled braiding $\tau_i$ and interchange $\tau_i^{(1)}$ to find
all three of the numbers \bea
|A_{01}|^2&\equiv& |\brak{\Psi}{n_1,\ldots,0,1,\dots, n_N}|^2,\nonumber\\
|A_{10}|^2&\equiv& |\brak{\Psi}{n_1,\ldots,1,0,\dots, n_N}|^2,\nonumber\\
|A_{10}+A_{10}|^2&\equiv& |\brak{\Psi}{n_1,\ldots,0,1,\dots,
n_N}+\brak{\Psi}{n_1,\ldots,1,0,\dots, n_N} |^2,\nonumber\\
|A_{01}+iA_{10}|^2 &\equiv&|\brak{\Psi}{n_1,\ldots,0,1,\dots,
n_N}+i\brak{\Psi}{n_1,\ldots,1,0,\dots, n_N} |^2\nonumber. \eea If
this is possible, we would be able to obtain all relative phases of
$\brak{\Psi}{n_1,\ldots,n_N}$'s, for given $|z_1|^2$, $|z_2|^2$,
$|z_1+z_2|^2$ and $|z_1+iz_2|^2$, it is possible to recover the
complex numbers $z_1$ and $z_2$ up to a common phase factor.

This is not quite the case, however. A vortex-antivortex interchange
would leave one vortex-vortex pair and one antivortex-antivortex
pair. These pairs are problematic, because the excited state of this
pair would not lead to an unbound Bogoliubov quasiparticle
excitation, but rather a bound excited state of a 'double
half-quantum' vortex (or antivortex). This means that fusing such
pairs would not give us a result that we can read out. In short,
$|z_1+iz_2|^2$ is not available to us.

Nevertheless, there is still much we can figure out about relative
phases of $\brak{\Psi}{n_1,\ldots,n_N}$'s. For $N$ complex numbers,
$z_i$, if one knows all their absolute values and distance between
each other, $|z_i - z_j|$, for all $i,j$ (or equivalently $|z_i +
z_j|$), the geometric configuration of $z_i$'s are determined {\it
rigidly}. By this we mean that we have determined $z_i$'s up to
overall rotation around the origin and reflection with respect to
the real axis (that is, complex conjugation). It needs to be noted
that knowing $|z_i \pm z_j|$ for all $i,j$ actually overdetermines
$z_i$'s. So it turns out that even though direct fusion combined
with controlled braiding $\tau_i$ only gives us a subset of $\{|z_i
\pm z_j|\}$, it is still sufficient for us to determine
$\brak{\Psi}{n_1,\ldots,n_N}$'s up to complex conjugation and
overall phase. The exact procedure will be given in the appendix.
(One needs to be note that if there is a superposition of even total
occupation number states and odd total occupation number states,
there is no way to obtain phase relation between them, which
reflects the fact that, as Ivanov pointed out \cite{ivanov}, the
superconducting Hamiltonian creates or destroys electrons only in
pairs.)

\section{Conclusions and open questions}

Half-quantum vortices in the $p_x+ip_y$ superconductors provide, at
least in principle, a way to generate and manipulate entangled
states in a topologically protected Hilbert space. We have shown
that by physically fusing vortices  with appropriate antivortices it
is possible to reconnect the  protected  space    to the rest of the
Hilbert space and so read out the information encoded there.

A number of questions remain, however: i)  We have assumed that the
fusion process is slow enough that the $E_i$ bound state merges
adiabatically with the continuum. What happens if the process is too
fast? Can we find a tractable model for the annihilation process
that would allow us to determine how adiabatic  it has to be in
order  not to lose information? ii)   Our picture of  generating and
moving vortices  is at present only a thought experiment. The
necessary half-quantum vortices have not even  been detected in any
real system.  If they can be found,  can   we come up with  some
practical device (a configuration of STM probes, current  sources,
drains, gates etc) that can create, guide, and monitor the vortices?
Since  no experiment is likely  to be able to detect a single
quasi-particle, we would have arrange for a continuous operation and
measure currents; iii) Perhaps the most interesting question is
whether  we take the insights developed from the relatively simple
picture of braiding and fusion in the superconductor and extend them
to the other candidate systems with  non-abelian statistics.

\section{Acknowledgments}
The early stages of this work were funded by the National Science
Foundation under grant NSF-DMR-01-32990. MS would like to thank
Eduardo Fradkin, Rinat Kedem and Eddy Ardonne for sharing their
insights into fusion rules and for useful conversations.

 \appendix

\section{Berry phases and  Bogoliubov transformations}

In this appendix we review the formal algebraic aspects of
Bogoliubov transformations, and their implications for the
computation of  Berry phases for BCS states.

\subsection{BCS ground state}

Suppose that $H_{ij}$ is an $N$-by-$N$ matrix representing a
one-particle Hamiltonian. When we include the effect of a
superconducting condensate, the second-quantized Bogoliubov
Hamiltonian becomes
\bea
 \hat H_{\rm Bogoliubov}&=& \ad_i H_{ij}a_j +\frac 12 \Delta_{ij} \ad_i\ad_j +\frac 12 \Delta^{\dagger}_{ij} a_i a_j\nonumber\\
&=& \frac12 \left(\matrix{ \ad_i &a_i}\right)\left(\matrix{ H_{ij}& \phantom {-}\Delta_{ij}\cr
                                                                              \Delta^{\dagger}_{ij}& -H^T_{ij}}\right)
                  \left(\matrix{ a_j\cr \ad_j}\right) +\frac 12 \tr H.
\eea
Here  $\ad_i$ and $a_i$ are fermion creation and annihilation
operators,the gap function $\Delta_{ij}$ is a skew symmetric matrix,
and $H^T$ denotes the transpose of the hermitian matrix $H$.  For a
continuum superconductor, the index  ``$i$''  should be understood
to incorporate both the space co-ordinate $x$, and the spin index. A
sum over $i$  therefore implies both an integral over real space and
a sum over spin components.

The many-body  Hamiltonian is diagonalized  by means of a
Bogoliubov transformation. To construct this transformation   we
begin by  solving  the single-particle  Bogoliubov-de-Gennes (BdG)
eigenvalue problem
\be
 \left(\matrix{ H & \Delta\cr
                        \Delta^{\dagger}& -H^T}\right)
                  \left(\matrix{ u_\alpha \cr  v_\alpha}\right)=E_\alpha  \left(\matrix{ u_\alpha \cr  v_\alpha}\right).
 \label{EQ:eigenvalue}
 \ee
Here  $u_\alpha$ and  $v_\alpha$ are $N$-dimensional column vectors,
which we take to be normalized so that
$|u_\alpha|^2+|v_\alpha|^2=1$. If we explicitly  display the
column-vector index $i$ they  become  matrices $u_{i\alpha}$ and
$v_{i \alpha }$. Taking the complex conjugate  of
(\ref{EQ:eigenvalue}) tells us  that
\be
 \left(\matrix{ H & \Delta\cr
                        \Delta^{\dagger}& -H^T}\right)
                  \left(\matrix{ v^*_\alpha \cr  u^*_\alpha}\right)=-E_\alpha  \left(\matrix{ v^*_\alpha \cr  u^*_\alpha}\right),
\label{EQ:negative_eigenvalue}
\ee
and so the BdG eigenvalues come
in $\pm$ pairs. We will   always take  $E_\alpha$ to be  the
positive eigenvalue.

 We now  set
 \bea
 a_i&=& u_{i\alpha}b_\alpha +v^*_{i\alpha}\bd_\alpha\nonumber\\
 \ad_i&=& v_{i\alpha} b_\alpha +u^*_{i\alpha}\bd_\alpha.
 \eea
The mutual orthonormality and completeness of the eigenvectors
$(u_\alpha, v_\alpha)^T$ ensures that the $b_\alpha$, $\bd_\alpha$
have the same anti-commutation relations as the $a_i$ $\ad_i$. In
terms of the $b_\alpha$ $\bd_\alpha$, the second-quantized
Hamiltonian becomes
\be
 \hat H_{\rm Bogoliubov} =\sum_{\alpha=1}^N E_\alpha \bd_\alpha b_\alpha -\frac 12 \sum_{\alpha=1}^N E_\alpha +\frac 12\sum_{i=1}^N E^{(0)}_i.
\label{EQ:bogdiag}
\ee
Here the $E^{(0)}_i$ are the eigenvalues of $H$. These can be of either sign.

If all the $E_\alpha$ are {\it strictly\/} positive, the new ground
state is non-degenerate and is the unique state $\ket{0}_b$
annihilated by all the $b_\alpha$. If we could find a unitary
operator $U$ such that \bea
b_\alpha &=& a_iu^*_{i\alpha}+ \ad_i v^*_{i\alpha}=  Ua_\alpha U^{-1}\nonumber\\
\bd_\alpha &=& \ad_iu_{i\alpha}+ a_i v_{i\alpha}=  U\ad_\alpha
U^{-1} \eea then we would have $\ket{0}_b=U\ket{0}_a $, where
$\ket{0}_a$ is the no-particle vacuum state. It is not easy to find
a closed-form expression for $U$, however. An alternative strategy
for obtaining $\ket{0}_b$  begins by  noting   that if   that the
matrix $u_{i\alpha}$ is invertible then the condition  $b_i
\ket{0}_b=0$  is equivalent to \be (a_i+\ad_k
v^*_{k\alpha}(u^{*-1})_{\alpha i})\ket{0}_b=0,  i=1,\ldots N. \ee We
therefore introduce  the skew-symmetric matrix \be S_{ij}=
v^*_{i\alpha}(u^*)^{-1}_{\alpha j} \ee and observe that \be
\exp\left\{\frac 12 \ad_i\ad_jS_{ij}\right\} a_k \exp\left\{-\frac
12 \ad_i\ad_jS_{ij}\right\}
 =a_k+\ad_iS_{ik}.
\ee
From this we conclude that
\be
 \ket{0}_b =N  \exp\left\{\frac 12 \ad_i\ad_jS_{ij}\right\} \ket{0}_a
 \label{EQ:BCS_vacuum}
 \ee
where $\ket{0}_a$ is the original-no particle state.  Equation
(\ref{EQ:BCS_vacuum}) explicitly  displays the superconducting
ground state  as a coherent  superposition of Cooper-pair states,
and allows us to identity $S_{ij}$ with the (unnormalized)  pair
wavefunction.

 The normalization factor ${\mathcal N}$ is found from
 \be
 \brak{S_1}{ S_2}= \det^{1/2}(I+S^\dagger_1 S_2),
 \ee
 where
\be
  \ket{S}= \exp\left\{\frac 12 \ad_i\ad_jS_{ij}\right\} \ket{0}_a,
 \ee
 to be
 \be
 {\mathcal N}= \det^{-1/4}(I+S^\dagger S).
\ee

Because the group of Bogoliubov transformations on $a_i$, $\ad_i$,
$i=1,\ldots, N$ is   ${\rm SO}(2N)$, and  because the subgroup
\hbox{${\rm U}(N)\simeq {\rm Sp}(2N, \mathbb R)\cap {\rm SO}(2N)$}
of  transformations that mix the $a$'s only with themselves ({\it
i.e.\/}\ not with the $\ad$'s), preserves the no-particle vacuum
$\ket{0}_a$,  the set of physically distinct   ground states is
parameterized by  the symmetric space  ${\rm SO}(2N)/{\rm U}(N)$. As
a check of this assertion, observe that \be {\rm dim\,}{\rm
SO}(2N)-{\rm dim\,}{\rm U}(N) = N(2N-1)-N^2=  N(N-1), \ee which is
the number of independent real parameters in the complex
skew-symmetric matrix $S_{ij}$. These  $S_{ij}$ for $i<j$  serve as
complex co-ordinates on all but a set of measure zero in the
manifold of possible ground states.

\subsection{Clifford  algebra and ${\rm Lie}({\rm SO}(2N))$}

We can make the ${\rm SO}(2N)$ character of the Bogoliubov
transformations manifest by introducing  a set of $2N$ Dirac gamma
operators. These are related to the fermion annihilation and
creation operators by
\bea
\gamma_i &=& (a_i+\ad_i)\nonumber\\
\gamma_{i+N} &=&  i (\ad_i-a_i). \eea The $\gamma_i$  are hermitian,
and obey the Clifford algebra \be \{\gamma_i,\gamma_j\}\equiv
\gamma_i\gamma_j+\gamma_j\gamma_i =2\delta_{ij}. \ee The Hamiltonian
can be rewritten in terms of the $\gamma_i$ as \be \hat H_{\rm
Bogoliubov}= \frac 12 \sum_{i,j=1}^{2N} h_{ij}\Gamma_{ij}+\frac 12
\tr H \ee where \be \Gamma_{ij}= \frac 1{4i}[\gamma_i,\gamma_j] \ee
are the spinor generators of the Lie algebra of ${\rm SO}(2N)$, and
the matrix $h_{ij}$ has entries \be h_{ij}=  \left(\matrix{ - \Im
H-\Im \Delta & -\Re H +\Re \Delta\cr
                                                             \phantom{-}     \Re H+\Re \Delta & -\Im H+\Im  \Delta}\right)_{ij}.
\ee Here $\Re Z \equiv X$ and $\Im  Z\equiv Y$ denote the real and
imaginary parts of $Z=X+iY$, and the vector space has been
partitioned so that $i=1,\ldots, N$ is the first block and
$i=N+1,\ldots 2N$ is the second. This rewriting reveals   that a
general Bogoliubov Hamiltonian is, up to an additive constant,  an
element of the Lie algebra  of ${\rm SO}(2N)$. The  Bogoliubov
transformation that diagonalizes $H$ is therefore the operation of
conjugating the Lie algebra  element $\hat H$  into the Cartan
sub-algebra. This  we can take to be spanned by the commuting set of
operators
\be
\frac 1{4i}[\gamma_{N+i},\gamma_i] = \frac{1}{2}(\ad_i
a_i- a_i\ad_i),\quad i=1,\ldots,N.
\ee

After conjugation,
\be
\hat H_{\rm Bogoliubov} \to U\hat H_{\rm Bogoliubov} U^{-1},
\ee
with
\be
U= \exp\left\{\frac{i}{2}\sum_{i,j} \theta_{ij} \Gamma_{ij}\right\}
\ee
for some parameters $\theta_{ij}$,
the Hamiltonian becomes.
\bea
\hat H_{\rm Bogoliubov}&\to &  \sum_{i=1}^N E_\alpha\frac 1{4i} [\gamma_{N+\alpha},\gamma_\alpha] +\frac 12 \sum_{i=1}^N E^{(0)}_i\nonumber\\
&=& \frac 12 \sum_{I=1}^N E_\alpha (\ad_\alpha a_\alpha-a_\alpha\ad_\alpha) +\frac 12 \sum_{i=1}^N E^{(0)}_i\nonumber\\
&=& \sum_{\alpha=1}^N E_\alpha \ad_\alpha a_\alpha  -\frac 12
\sum_\alpha E_\alpha+ \frac 12\sum_i E^{(0)}_i ,
\eea
which is the
same as (\ref{EQ:bogdiag}). Again it is convenient to regard  the
energies  $E_\alpha$ as being positive, even though the $E^{0}_i$,
the eigenvalues of $H$, can have either sign. The additive constant
is then the ground-state energy of superconducting system.

\subsection{Zero modes}

The gamma-operator language is particularly useful when there are
zero modes. Because of the $\pm E$ symmetry, any zero energy
eigenvectors of the BdG Hamiltonian  must come in pairs. Suppose
that $(u_0,v_0)^T$ becomes degenerate with its negative energy
sibling $(v^*_0, u^*_0)^T$. Then we can write this pair of
eigenvectors'  contribution to the mode expansion as
\bea
\left(\matrix{a_i\cr \ad_i}\right)&=&\left(\matrix{u_{i0} \cr
v_{i0}}\right)b_0 +
\left(\matrix{v^*_{i0} \cr u^*_{i0}}\right) \bd_0+\cdots\nonumber\\
 &=&
\left(\matrix{U_{i0} \cr V_{i0}}\right) \frac {\gamma_0}{\sqrt{2}} +
\left(\matrix{U_{iN} \cr V_{iN}}\right)   \frac {\gamma_N}{\sqrt{2}}
+\cdots,
\eea
where
\be
\left(\matrix{U_{i0} \cr V_{i0}}\right)
=\frac 1{\sqrt{2}}\left[\left(\matrix{u_{i0} \cr v_{i0}}\right) +
\left(\matrix{v^*_{i0} \cr u^*_{i0}}\right)\right],\quad
\left(\matrix{U_{iN} \cr V_{iN}}\right) =\frac
i{\sqrt{2}}\left[\left(\matrix{u_{i0} \cr v_{i0}}\right)
-\left(\matrix{v^*_{i0} \cr u^*_{i0}}\right)\right],
\ee
and
\be
\gamma_0=(b_0+\bd_0),\quad \gamma_N=i(\bd_0-b_0).
\ee
The column
vectors $(U_0,V_0)^T$ and $(U_N,V_N)^T$ both have the feature that
$U^*=V$. This anti-linear up-down  symmetry is characteristic of the
localized zero modes in the vortex cores.

\subsection{Berry Connection}

We need to compute the Berry connection $ iA= \eval{\tilde
S}{d}{\tilde S} $, where $\ket{\tilde S}={\mathcal N}\ket{S}$ is the
normalized ground state. To do this we exploit the fact that  the
un-normalized states $\ket{S}$,  being  functions only of the
$S_{ij}$, and not of the  $S_{ij}^*$, defines  a holomorphic
line-bundle over the K\"ahler manifold   ${\rm SO}(2N)/{\rm U}(N)$.
We can therefore read off the  1-form connection $iA$ from
derivatives of the  K\"ahler potential
\be
 \ln {\mathcal N}= -\frac 14 \ln \det (I+S^\dagger S)
\ee
as $iA=\bar\partial \ln {\mathcal N}-\partial \ln {\mathcal N}$
\cite{stone_QH_book}. If we express the parameters $S_{ij}$ in terms
of the normalized Bogoliubov eigenvectors $(u,v)^T$, we find that
\bea
iA&=&  \sum_{i<j} \left(\frac{\partial \ln {\mathcal N}}{\partial S^*_{ij}}   \, dS^*_{ij}  - \frac{\partial \ln {\mathcal N}}{\partial S_{ij}} \, dS_{ij}\right)\nonumber\\
 &=& \frac 12 \sum_{\alpha=1}^N \left(\matrix{v_\alpha & u_\alpha} \right) d \left(\matrix{ v^*_\alpha \cr u^*_\alpha}\right) +\frac{i}{2} d \left\{ {\rm Arg\,} ( \det u) \right\}.
\label{EQ:majorana_berry}
\eea
The expression
(\ref{EQ:majorana_berry}) has a simple interpretation. From
(\ref{EQ:negative_eigenvalue}) we see that the column vectors
$(v^*_\alpha, u^*_\alpha)^T$ are the negative-energy eigenstates  of
the one-particle Bogoliubov-de-Gennes Hamiltonian
\be
H_{\rm BdG}=
\left(\matrix{ H & \Delta\cr
                        \Delta^{\dagger}& -H^T}\right).
 \ee
If this were a Dirac fermion problem, we would fill the Dirac sea
consisting  of these negative energy states. The Berry phase of the
vacuum would then be the sum of the Berry phases of the occupied
states. The first term in  (\ref{EQ:majorana_berry}) is precisely
one-half of this sum. The factor of one-half compensates for  the
artificial doubling of the degrees of freedom in passing from $H$ to
$H_{\rm BdG}$. The second term in (\ref{EQ:majorana_berry})  is a
total derivative, and reflects a choice of gauge.

We next compute the Berry connection for an excited state
\be
\ket{\alpha_1,\ldots,\alpha_n}= \bd_{\alpha_1}\cdots\bd_{\alpha_n}\ket{\tilde S},
\ee
by using
\bea
d \bd_\alpha &=&(u_{i\beta}b_{\beta} +v^*_{i\beta}\bd_\beta)\,dv_{i\alpha}+(v_{i\beta}b_{\beta}+
u^*_{i\beta}\bd_\beta)\,du_{i\alpha}\nonumber\\
&=&
(v^*_{i\beta}dv_{i\alpha}+u^*_{i\beta}du_{i\alpha})\,\bd_{\beta}+
(u_{i\beta}dv_{i\alpha}+v_{i\beta}du_{i\alpha})\,b_{\beta}.
\eea
When the state $\ket{\alpha_1,\ldots,\alpha_n}$ is non-degenerate,
we are interested only in the diagonal $\beta=\alpha$ term
\be
d\bd_\alpha|_{\rm diag}=
(v^*_{i\alpha}dv_{i\alpha}+u^*_{i\alpha}du_{i\alpha} )\,\bd_\alpha,
\quad \hbox{(no sum on $\alpha$)} ,
\ee
and we find
\bea
iA &=& \eval{\alpha_1,\ldots,\alpha_n}{d}{\alpha_1,\ldots,\alpha_n}\nonumber\\
&=&\eval{\tilde S}{b_{\alpha_n}\cdots b_{\alpha_1} d\left( \bd_{\alpha_1}\cdots\bd_{\alpha_n}\right.}{\tilde S}\left.\!\!\right)\nonumber \\
  &=& \eval{\tilde S}{d}{\tilde S} + \eval{\tilde S}{b_{\alpha_n}\cdots b_{\alpha_1} d\left( \bd_{\alpha_1}\cdots\bd_{\alpha_n}\right)\!}{\tilde S}\nonumber\\
  &=&  \eval{\tilde S}{d}{\tilde S}  + \sum_{m=1}^{n} \left(\matrix {u^*_{\alpha_m} & v^*_{\alpha_m}}\right)
  d \left(\matrix{u_{\alpha_m}\cr v_{\alpha_m}}\right),
\eea
which is the sum of the many-body ground-state Berry connection
and the Berry connections of the individual  one-particle  excited
states. Observe that there  is  no factor of $``1/2$'' in  the
contribution of these occupied excited states.

The non-abelian Berry connection of a set of degenerate many -body
states is computed in the same manner. It will include  a diagonal
term from the reference state $\ket{\tilde S}$ and a sum of
non-diagonal terms terms of the form
\be
iA_{\beta\alpha} =
\left(\matrix{u^*_\beta & v^*_\beta} \right) d \left(\matrix{
u_\alpha \cr v_\alpha}\right).
\ee
In the case of the vortices, the
states of interest are the exponentially localized core states.
Because of this localization only the overlap of each core state
with itself has any chance of providing a non-zero term in the
connection, but it is readily verified that with the phase choices
made in the text, all these contributions are zero. The only ``Berry
phase'' produced by the vortex braiding is the overall diagonal
Berry phase associated with the Magnus force \cite{ao}

\section{Determining phase relation among coefficients of occupation number basis}

Let us consider how many vortex interchange steps would be needed in
order to {\it rigidly} - this rigidity being defined in the last
page of the section III - determine geometric configuration on the
complex plane of all $\brak{\Psi}{n_1,\ldots,n_N}$'s, with
$|\Psi\rangle$ having a definite parity in total occupation number.

In the case $N=2$ one can easily see from \bea \tau_1
(A_{00}|00\rangle +A_{11}|11\rangle) &=& \frac{1}{\sqrt{2}}\{(A_{00}
+
A_{11})|00\rangle - (A_{00} - A_{11})|11\rangle \},\nonumber\\
\tau_1 (A_{01}|01\rangle +A_{10}|10\rangle) &=&
\frac{1}{\sqrt{2}}\{(A_{01} + A_{10})|01\rangle - (A_{01} -
A_{10})|10\rangle\} \eea that the fusion following the vortex
interchange process gives us, up to the sign, phase difference
between $A_{00}$  and $A_{11}$ (and likewise between $A_{01}$ and
$A_{10}$). Given that we already know $|A_{00}|$ and $|A_{11}|$,
this is sufficient to rigidly determine on the complex plane the
configuration of $A_{00}$  and $A_{11}$. Same can be said for
$A_{01}$ and $A_{10}$.

It is instructive to work out the next simplest case $N=3$, where
$\ket{\Psi} = A_{000}|000\rangle +A_{110}|110\rangle +
A_{011}|011\rangle +A_{101}|101\rangle$. From $N=2$ case, one can
see that $\tau_1$ rigidly determines the configuration of $A_{000}$
and $A_{110}$ on the complex plane. It also rigidly determines the
configuration  of $A_{011}$ and $A_{101}$ on the complex plane.
However any phase relation between the former and latter remains
completely unknown at this point. But then from \be \tau_2
\ket{\Psi} = \frac{1}{\sqrt{2}} \{(A_{000}+A_{011})|000\rangle
-(A_{101}-A_{110})|110\rangle - (A_{000}-A_{011})|011\rangle
+(A_{101} +A_{110})|101\rangle\},\ee one can see that after fusion
following the implementation of $\tau_2$, the configuration of both
the $A_{000}$, $A_{011}$ pair and the $A_{101}$, $A_{110}$ pair
would be determined rigidly with respect to the origin. We now have
a rigid configuration of all four coefficients - $A_{000}$,
$A_{011}$, $A_{101}$, and $A_{110}$ - on the complex plane; they
have been determined up to complex conjugation and overall phase.

The $N=3$ case gives us ideas about how to make use of a recursion
argument for the general case in figuring out the phase relation.
One can assume that for $N = m$ there is some process consisting of
fusion of vortex-antivortex pairs and applications of $\tau_k$'s
(where $k \leq m-1$) which give us the rigid configuration on the
complex plane of the coefficient of states with even total
occupation number. The same process would also give us the rigid
configuration on the complex plane of the coefficient of states with
odd total occupation number as well. (We have seen that this holds
true for $m=2$.) Now note that the even (or odd) total occupation
number sector of the Hilbert space in $N=m+1$ case can be divided
into the following two classes: \bea \label{EQ:class1}
|n_1,\ldots,n_m,0\rangle,\\
\label{EQ:class2} |n'_1,\ldots,n'_m,1\rangle. \eea The operations
that were used in the procedure we have applied for obtaining all
phase relation in the $N=m$ case does not affect $n_{m+1}$, and so
by applying these operations we would obtain the rigid configuration
for the coefficients of states belonging to (\ref{EQ:class1}). Same
can be said for the coefficients of the states of belonging to
(\ref{EQ:class2}) (though $\ket{n_1,\ldots,n_m}$ and
$\ket{n'_1,\ldots,n'_m}$ have opposite parity in total occupation
number). Now all we need to do is to figure out the rigid
configuration of two pairs of coefficients, each of which consists
of one coefficient for one of the states belonging to
(\ref{EQ:class1}) and one coefficient for one of the states
belonging to (\ref{EQ:class2}). So for $|\Psi\rangle = \sum
A_{n_1,\ldots,n_m,n_{m+1}}\ket{n_1,\ldots,n_m,n_{m+1}}$, (note that
summation is restricted to even, or odd, total occupation number)
\bea \tau_m |\Psi\rangle &=& \tau_m \left( \sum A_{n_1,\ldots,n_m,0}
\ket{n_1,\ldots,n_m,0}+ \sum A_{n'_1,\ldots,n'_m,1}
\ket{n'_1,\ldots,n'_m,1}\right)
\nonumber\\
&=& \frac{1}{\sqrt 2} \{\sum (A_{n_1,\ldots,n_{m-1},0,0} +
A_{n_1,\ldots,n_{m-1},1,1}) \ket{n_1,\ldots,n_{m-1},0,0}\nonumber\\
&\,& \,\,\,\,\,\,\,\,\,\,\,\,\,\,\, - \, (A_{n_1,\ldots,n_{m-1},0,0}
- A_{n_1,\ldots,n_{m-1},1,1})
\ket{n_1,\ldots,n_{m-1},1,1}\}\nonumber\\
&+& \,\frac{1}{\sqrt 2} \{\sum (A_{n'_1,\ldots,n'_{m-1},0,1}
+A_{n'_1,\ldots,n'_{m-1},1,0})\ket{n'_1,\ldots,n'_{m-1},0,1}\nonumber\\
&\,& \,\,\,\,\,\,\,\,\,\,\,\,\,\,\,\, - \,
(A_{n'_1,\ldots,n'_{m-1},0,1}
-A_{n'_1,\ldots,n'_{m-1},1,0})\ket{n'_1,\ldots,n'_{m-1},1,0}\}. \eea
One can easily see that all pairs belonging to
($A_{n_1,\ldots,n_{m-1},0,0}$, $A_{n_1,\ldots,n_{m-1},1,1}$) or
($A_{n'_1,\ldots,n'_{m-1},0,1}$, $A_{n'_1,\ldots,n'_{m-1},1,0}$) now
have rigid configuration. It is clear that we now have rigid
configuration for $\{A_{n_1,\ldots,n_{m-1},n_m,n_{m+1}}\}$. Also we
can see that in this scheme of figuring out the rigid configuration
of $2^{N-1}$ coefficients of even (or odd) fermion sector in the
occupation-number basis, $N-1$ vortex interchange steps are needed.

\end{document}